\documentclass[aps,superscriptaddress,floatfix,showpacs,letterpaper,nofootinbib]{revtex4}

\usepackage{graphicx}
\usepackage{enumerate}

\def\bwt{\begin{widetext}}
\def\ewt{\end{widetext}}
\def\be{\begin{equation}}
\def\ee{\end{equation}}
\def\bea{\begin{eqnarray}}
\def\eea{\end{eqnarray}}
\def\bean{\begin{eqnarray*}}
\def\eean{\end{eqnarray*}}
\def\bary{\begin{array}}
\def\eary{\end{array}}
\def\bit{\begin{itemize}}
\def\eit{\end{itemize}}

\def\su5u1{SU(5) \times U(1)}
\def\fsu5u1{SU(5) \times U(1)'}
\def\so10{SO(10)}
\def\sq20{SO(10) \times SO(10)}

\begin{document}

\title{Effects of Dimension-5 Operators in $E_6$ Grand Unified Theories}

\author{Chao-Shang Huang}\email{csh@itp.ac.cn}
\affiliation{State Key Laboratory of Theoretical Physics,
 Institute of Theoretical Physics, Chinese Academy of Sciences,
Beijing 100190, P. R. China}


\date{\today}

\begin{abstract}

The effective dimension-5 operators can be induced by quantum gravity or inspired by string and M theories.
They have important impacts on grand unified theories. We investigate 
their effects for the well known E(6) model. Considering the breaking chains $E_{6}\mapsto H=SO(10)\times U_{V'}(1)\mapsto SU(5)\times U_{V}(1)\times U_{V'}(1)\mapsto SU(3)\times SU(2)\times U_{Z}(1)\times U_{V}(1)\times U_{V'}(1)$ and $E_{6}\mapsto H=SO(10)\times U_{V'}(1)\mapsto SU(4)\times SU_{L}(2)\times SU_{R}(2)\times U_{V'}(1)\mapsto SU(3)\times SU_{L}(2)\times SU_{R}(2)\times U_{S}(1)\times U_{V'}(1)$, we derive and give all of the Clebsch-Gordan coefficients $\Phi^{(r)}_{s,z}$ associated with $E_6$ breaking to the standard model. Physical effects of nonzero vacuum expectations of SM singlets Higgs in $E_6$ grand unified theories are discussed.

\end{abstract}

Keywords: Dimension-5 operators $E_6$ GUT  

\pacs{12.10.Dm, 12.10.Kt, 04.60.-m}


\maketitle

Grand unification theories (GUT) are among the most promising models
for physics beyond the standard model (SM). Grand unification assumes that the
three gauge coupling constants in SM should be unified  at a high
scale, the unification scale $M_G$, but would be split at low energy due to their different renormalization
group evolution from the grand unification scale to low scales. It seems that the assumption is supported by
experiments since there is the apparent
unification of the measured gauge couplings within the minimal SUSY SM
(MSSM) at scale $M_{G}\sim 2\times 10^{16}$ GeV \cite{a,a1,a2,a3}.
In addition to the GUT scale, one has the Planck scale defined by
$ M_{Pl}=(8\pi G_{N})^{-1/2}\sim 2.4\times 10^{18}$ GeV at which
quantum gravity enters \cite{npq,t}. It is well-known that
the GUT scale $M_{G}$ is smaller than the
Planck scale $ M_{Pl}$ by two orders of magnitude. Therefore, we can investigate
unification of particle interactions including effects of gravity in the effective field theory approach. That is,
we introduce non-renormalizable higher dimension operators to describe the effects of quantum gravity.
They are $d\geq 5$ operators which are induced by gravity, enter the Lagrangian scaled by factors
Of $(M_{Pl})^{-(d-4)}$ with order unity coefficients, and are subject
only to the constraints of the symmetries (gauge invariance,
supersymmetry, etc.) of the low energy theory.
The presence of higher-dimensional operators generated at the Planck scale must have impact to GUT and its phenomenology, as has been shown in Refs. ~\cite{gc,gc1,gc2,gc3,gc4,gc5,gc6,pd,npq,ywz,sp,sp1,sp2,sp3,sp4,sp5,dmn,chlw,chlw1}. These operators modify the usual gauge coupling unification condition~\cite{gc,gc1,gc2,gc3,gc4,gc5,gc6}. They affect analysis of proton decay~\cite{pd,npq,ywz} and supersymmetric (SUSY) particle spectrum in SUSY GUT and supergravity~\cite{sp,sp1,sp2,sp3,sp4,sp5,chlw,chlw1}. Therefore, one should consider effects of these higher dimension operators in model building of GUT and SUSY GUT.

The effects to the unification of gauge couplings produced by dimension-5 operators which are singlets of the grand unified gauge group G and formed from gauge field strengths $G_{\mu\nu}$ and Higgs multiplets $H_{k}$ of G
\be
{\cal L} =\frac{c_{k}}{M_{Pl}} G^{a}_{\mu\nu} G^{b \mu\nu} H_{k}^{ab}, \label{dim5}
\ee
where a,b are group indices and k labels different multiplets, have been examined systematically for $G = SU(5), SO(10)$ in the ref.~\cite{chr}. In particular, differing from those given before in the literature, all of the Clebsch-Gordan coefficients $\Phi^s_{(r)t}$ associated with $SU(5)$ and $SO(10)$ breaking to the standard model, in different bases $\{t\}$, have been derived and given, upon a uniform absolute normalization scheme across different representations $r$, in the reference.

It is well known that the exceptional group $E_{6}$ is an attractive unification group. From the viewpoint of superstring theory, the gauge and gravitational anomaly cancelation occurs only for the gauge
groups SO(32) or $E_{8} \times E_{8}$~\cite{gs,gs1} and compactification on a Calabi-Yau manifold
with an SU(3) holonomy results in the breaking $E_{8} \rightarrow SU(3) \times E_{6}$.
This fact inspired the interests in E6 GUT. Furthermore, the dynamical symmetry breaking scenario would give several constraints on the possible GUT models. It has been pointed that
$E_{6}$ is uniquely selected among many GUT groups if we require 1) every generation of quarks/lepton
fields belongs to a single irreducible representation (irrep) of the GUT group, 2) the theory is
automatically anomaly free, and 3) all the Higgs fields, which are necessary for causing the symmetry breaking down to
$SU_{c}(3)\times U_{em}(1)$, fall in the representations that can be provided by the fermion bilinears~\cite{bn,bn1,ram}.

In the letter we investigate effects of the operators, Eq. (\ref{dim5}), for $G = E_6$. We derive and give all of the Clebsch-Gordan coefficients $\Phi^{(r)}_{s,z}$ associated with $E_6$ breaking to the standard model, in different bases $\{s,z\}$, upon a uniform absolute normalization for different representations $r$. The detailed analysis of RG evolution of the gauge couplings, when the operators, Eq. (\ref{dim5}), exist in the effective Lagrangian, is not presented in this letter and leaves in the future.

It is evident from Eq. (\ref{dim5}) that the representations to which Higgs fields $H_{k}$ belong can only be contained in the symmetric product of two adjoints \footnote{For simplicity, we use "$\times$" and "+" to denote the direct product and the direct sum respectively whenever there is no confusion.}
\begin{eqnarray}
({\bf 78}\times{\bf 78})_{symmetric}&=&{\bf 1}+{\bf 650}+{\bf 2430}. \label{reps}
\end{eqnarray}
For a specific irrep $r,~r=1,650,2430$, in Eq. (\ref{reps}), we denote the Higgs multiplet by a d-dimensional symmetric matrix
$\Phi^{(r)}$ with d=d(G), the dimension of the adjoint representation (rep) G ( we use the same letter G to denote the group and its adj. rep. for simplicity ) and d=78 for $G=E_6$. For our purpose we find the
all possible $\Phi^{(r)}$ which are invariant under the standard model gauge group $G_{321}\equiv SU(3)\times SU(2)\times U_{Y}(1)$. That is, each of them is a SM singlet and in this case the matrix is largely simplified: it contains only a few independent entries.
Looking at the branching rule for the GUT group $E_{6}$ \cite{slansky81}, we see that there are several maximal subgroups which contain $G_{321}$ (e.g., $H=SO(10)\times U(1), H=SU(3)\times SU(3)\times SU(3),H=SU(2)\times SU(6), H=F_{4}$). For a specific maximal subgroup, there are several breaking chains usually. For example, for $H=SO(10)\times U(1)$ there are two well-known breaking chains, (1) $E_{6}\mapsto H=SO(10)\times U_{V'}(1)\mapsto SU(5)\times U_{V}(1)\times U_{V'}(1)\mapsto SU(3)\times SU(2)\times U_{Z}(1)\times U_{V}(1)\times U_{V'}(1)$ and (2) $E_{6}\mapsto H=SO(10)\times U_{V'}(1)\mapsto SU(4)\times SU_{L}(2)\times SU_{R}(2)\times U_{V'}(1)\mapsto SU(3)\times SU_{L}(2)\times SU_{R}(2)\times U_{S}(1)\times U_{V'}(1)$.  We consider the case of $H=SO(10)\times U(1)$ in the letter and leave the study for cases of other maximal subgroups in $E_{6}$ as well as related breaking chains in the future.

For the breaking chain (1), there are three ways to define the hypercharge $Y$ which are consistent with the SM ~\cite{sy,sy1}: 1) Y/2=Z, i.e., $ SM \subset SU(5)$; 2) Y/2=-(Z+V)/5, i.e., $SM \subset SU(5)\times U_{V}(1)$; 3) Y/2=-(4 Z-V-5 V')/20, i.e., $SM \subset SU(5)\times U_{V}(1)\times U_{V'}(1)$. In the cases 1) and 2), $U_{Y}(1)$ is a subgroup of $SO(10)$, of which we shall call the case 1) "normal embedding" hereafter, and in the case 3) $U_{Y}(1)$ is a subgroup of $E_{6}$, which we shall call "flipped embedding".
\begin{enumerate}
\item normal embedding $G_{321}\subset SU(5)\subset SO(10)\subset E_6$
\begin{eqnarray}\label{e6nomal}
{\bf 78} &\stackrel{SO(10)\times U_{V'}(1)}{\longrightarrow}& {\bf 45_0}\oplus{\bf 1_0}~~\oplus~~\left({\bf 16_{-3}}\oplus {\rm h.c.}\right)\nonumber\\
&\stackrel{SU(5)\times U_{V}(1)\times U_{V'}(1)}{\longrightarrow}& {\bf 24_{0,0}}\oplus{\bf 1_{0,0}}~~\oplus~~\left({\bf 10_{4,0}}\oplus {\rm h.c.}\right)\oplus{\bf 1_{0,0}}
 \nonumber\\ && \oplus ~~\left({\bf 10_{-1,-3}}\oplus \overline{\bf {5}}_{3,-3} \oplus {\bf 1_{-5,-3}} \oplus {\rm h.c.}\right) \nonumber\\
&\stackrel{G_{321}}{\longrightarrow}& \underbrace{({\bf 8},{\bf 1})_0}_{I} \oplus \underbrace{({\bf 1},{\bf 3})_0}_{II} \oplus \underbrace{({\bf 1},{\bf 1})_0}_{III} \oplus \underbrace{\left(({\bf 3},{\bf 2})_{-\frac{5}{6}} \oplus {\rm h.c.}\right)}_{IV}\oplus\underbrace{({\bf 1},{\bf 1})_0}_{V}\nonumber\\
&&~~\oplus~~\underbrace{\left(({\bf 3},{\bf 2})_{\frac{1}{6}} \oplus {\rm h.c.}\right)}_{VI} \oplus \underbrace{\left((\overline{\bf 3},{\bf 1})_{-\frac{2}{3}} \oplus {\rm h.c.}\right)}_{VII} \oplus \underbrace{\left(({\bf 1},{\bf 1})_1 \oplus {\rm h.c.}\right)}_{VIII} \oplus~~\underbrace{({\bf 1},{\bf 1})_0}_{IX} \nonumber\\
&&~~\oplus \underbrace{\left(({\bf 3},{\bf 2})_{\frac{1}{6}}\oplus {\rm h.c.}\right)}_{X}\oplus \underbrace{\left((\overline{\bf 3},{\bf 1})_{-\frac{2}{3}} \oplus {\rm h.c.}\right)}_{XI}
 \oplus \underbrace{\left(({\bf 1},{\bf 1})_{1} \oplus {\rm h.c.}\right)}_{XII} \nonumber\\
 &&~~\oplus  \underbrace{\left((\overline{\bf 3},{\bf 1})_{\frac{1}{3}} \oplus {\rm h.c.}\right)}_{XIII}\oplus \underbrace{\left(({\bf 1},{\bf 2})_{-\frac{1}{2}} \oplus {\rm h.c.}\right)}_{XIV}\oplus \underbrace{\left(({\bf 1},{\bf 1})_{0} \oplus {\rm h.c.}\right)}_{XV}.
\end{eqnarray}
\item flipped embedding $G_{321}\subset SO(10)\times U_{V'}(1)\subset E_6 $
\begin{eqnarray}\label{decompSO10flipped}
{\bf 78} &\stackrel{SO(10)\times U_{V'}(1)}{\longrightarrow}& {\bf 45_0}\oplus{\bf 1_0}~~\oplus~~\left({\bf 16_{-3}}\oplus {\rm h.c.}\right)\nonumber\\
&\stackrel{SU(5)\times U_{V}(1)\times U_{V'}(1)}{\longrightarrow}& {\bf 24_{0,0}}\oplus{\bf 1_{0,0}}~~\oplus~~\left({\bf 10_{4,0}}\oplus {\rm h.c.}\right)\oplus{\bf 1_{0,0}}
 \nonumber\\ && \oplus ~~\left({\bf 10_{-1,-3}}\oplus \overline{\bf {5}}_{3,-3} \oplus {\bf 1_{-5,-3}} \oplus {\rm h.c.}\right) \nonumber\\
&\stackrel{G_{321}}{\longrightarrow}& \underbrace{({\bf 8},{\bf 1})_0}_{I} \oplus \underbrace{({\bf 1},{\bf 3})_0}_{II} \oplus \underbrace{({\bf 1},{\bf 1})_0}_{III'} \oplus \underbrace{\left(({\bf 3},{\bf 2})_{\frac{1}{6}} \oplus {\rm h.c.}\right)}_{IV}\oplus\underbrace{({\bf 1},{\bf 1})_0}_{V'}\nonumber\\
&&~~\oplus~~\underbrace{\left(({\bf 3},{\bf 2})_{\frac{1}{6}} \oplus {\rm h.c.}\right)}_{VI} \oplus \underbrace{\left((\overline{\bf 3},{\bf 1})_{\frac{1}{3}} \oplus {\rm h.c.}\right)}_{VII} \oplus \underbrace{\left(({\bf 1},{\bf 1})_0 \oplus {\rm h.c.}\right)}_{VIII} \oplus~~\underbrace{({\bf 1},{\bf 1})_0}_{IX'} \nonumber\\
&&~~\oplus \underbrace{\left(({\bf 3},{\bf 2})_{-\frac{5}{6}}\oplus {\rm h.c.}\right)}_{X}\oplus \underbrace{\left((\overline{\bf 3},{\bf 1})_{-\frac{2}{3}} \oplus {\rm h.c.}\right)}_{XI}
 \oplus \underbrace{\left(({\bf 1},{\bf 1})_{-1} \oplus {\rm h.c.}\right)}_{XII} \nonumber\\
 &&~~\oplus  \underbrace{\left((\overline{\bf 3},{\bf 1})_{-\frac{2}{3}} \oplus {\rm h.c.}\right)}_{XIII}\oplus \underbrace{\left(({\bf 1},{\bf 2})_{-\frac{1}{2}} \oplus {\rm h.c.}\right)}_{XIV}\oplus \underbrace{\left(({\bf 1},{\bf 1})_{-1} \oplus {\rm h.c.}\right)}_{XV}.
\end{eqnarray}
\end{enumerate}
Therefore, $\Phi^{(r)}$ which are invariant under $G_{321}$, can be written as
\bea  \label{phir}
(\Phi^{(r)})^{ab}&=& \delta^{ab}~diag[h1~I_8,h2~I_3,h3,h4~I_{12},h5,h6~I_{12},h7~I_6,h8~I_2,h9,h10~I_{12},h11~I_6,h12~I_2,\nonumber\\
&&~~h13~I_6,h14~I_4,h15~I_2]+\delta^{a12}\delta^{b25} h35+\delta^{a~25}\delta^{b~12} h53+\delta^{a~12}\delta^{b~46}h39\nonumber\\
&&~~+\delta^{a~25}\delta^{b~46} h59+\delta^{a~46}\delta^{b~12} h93+\delta^{a~46}\delta^{b~25} h95 .
\eea
Where $I_n$ is n-dimensional unit matrix, $hij=hji$ since $\Phi^{(r)}$ is the symmetric matrix and we have ordered indices a,b according to the order of terms in Eq.(\ref{e6nomal}), i.e., a=1,...,8 for h1, a=9,10,11 for h2, a=12 for h3, a=13,...,24 for h4, ..., etc. That is, we have chosen such classification of generators $X^a$ that they are identified according to their transformation properties under subgroups of $E_6$. Because $hj$, j=1,2,...,15, are the same for both the normal and flipped embedding except for $h3, h5, h9$, we denote the different entries in the case of flipped embedding by $h3', h5', h9'$ and $hij'$.

We now come to a comment. Besides the generators III, V, IX, there are some terms in eq. (\ref {e6nomal}) which have same SM quantum numbers. For example, XV with III, V and IX, VI with X, etc. These terms have no mixing with the other terms of same SM quantum numbers, i.e., corresponding $hij$ =0. The reason is as follows. For the breaking chain (1), we have $E_{6}\mapsto SU_{C}(3)\times SU_{L}(2)\times U_{Z}(1)\times U_{V}(1)\times U_{V'}(1)$ finally. The different assignments of hypercharge give the different mixing among three $U(1)$'s so that only $hij$, i,j=3,5,9, can be non-vanishing since i=3,5,9 correspond to the generators $U_{Z}(1), U_{V}(1), U_{V'}(1)$ respectively. Similar discussions are also valid for Eq. (\ref{decompSO10flipped}). 

In order to find $\Phi^{(r)}$ we use the second order Casimir operator
\be \label{casimir}
C_R=-\sum_{i=1}^{d(G)} X_i^2, \ee
where R is a irrep of G and $X_i$'s are the generators of G which satisfy
\be
[X_i,X_j]= i~f_{ijk} X_k \ee
with $f_{ijk}$ being the totally antisymmetric structure constants. The operator acts on the tensor product $R\times R$ so that
\be
C_{R\times R}= C_R\times {\bf {\it 1}} + {\bf {\it 1}} \times C_R + 2 F, \ee where
\be \label{ff}
F= -\sum_{i=1}^{d(G)} X_i \times X_i . \ee
It is easy to derive $F \Phi^{(r)}= F_{G}(r) \Phi^{(r)}$, where $F_{G}(r)=C(r)/2-C(G) $ is the eigenvalue of $F$
in irrep r, $C(r)$ and $C(G)$ the eigenvalues of the Casimir operator in irrep r and G respectively. $C(r)$ depends the normalization of the Casimir operator and consequently the choice of structure constants. The structure constants of $E_6$ have been given in a Chevalley base~\cite{sc}. We transform them into the usual form\footnote{Because to list them needs a long place, we will present them in a regular paper in the future.}, which are mostly used by physicists and in study of GUT, and choose them such that the part corresponding to $SO(10)$ is the same as that in the ref.~\cite{chlw,chr,gss}. The eigenvalues $C(r)$ for several irreps r in $E_6$ as well as $SO(10)$ are listed in Table I.

Using Eq.(\ref{ff}), a straightforward but tedious computation leads our results for $E_6$. The results for $\Phi^{(r)}$ \footnote{We normalize each $\Phi^{r}_{s,z}$ such that $Tr(\Phi^{r}_{s,z} \Phi^{r}_{s,z})=1$, see Table II and Table IV.} classified by the subgroup in the breaking chain are shown in Table \ref{su5} and \ref{su5n}. And the results for the breaking chain (2) are shown in Table \ref{su4} and \ref{su4n}. In the Tables \ref{su5} and \ref{su5n}, for the $r={\bf 1}$, s=1, z=1; for r={\bf 650}, s=1,2,3,4 correspond to {\bf 1}, {\bf 45}, {\bf 54} and {\bf 210} of $SO(10)$ respectively and z=1 for {\bf 1} of $SO(10)$, z=1,2 correspond to {\bf 1}, {\bf 24} of $SU(5)$ respectively when s=2, z=1 when s=3, z=1,2,3 correspond to {\bf 1}, {\bf 24}, {\bf 75} of $SU(5)$ respectively when s=4; similarly for r={\bf 2430}. The same as for the case of $G=SO(10)$~\cite{chr}, such a classification can serve as a parametrization for SM singlets $ <H_k^{ab}> $, the nonzero vacuum expectations of $H_k^{ab}$, transforming in irrep $r_k$ \footnote{The convention is: k=1,2,3 correspond to irreps ${\bf 1},{\bf 650},{\bf 2430}$.}   of $E_6$, in terms of the basis $\Phi^{k}_{s,z}\equiv \Phi^{(r_k)}_{s,z}$:
\be \label{vavs}
<H_k^{ab}> \equiv \sum_{s,z} v_{s,z}^{k}~ \Phi^{k~ab}_{s,z}\equiv v^k~\Phi^{k~ ab}, \ee
where $v_{s,z}^{k}, v^k$ are real.
A direct physical effect of Eqs. (\ref{dim5}),(\ref{vavs}) is that they lead to an alteration of the gauge coupling unification
condition:
\be \label{gaucon}
g^2_1(M_{GUT})(1 + \epsilon_1) = g^2_2(M_{GUT})(1 + \epsilon_2) = g^2_3(M_{GUT})(1 + \epsilon_3)=g^2_G/4\pi, \ee
where the
\be
\epsilon_i = \sum_k \frac{c_k}{M_{Pl}} \sum_{s,z} v_{s,z}^k \phi_i^{k~sz}, ~~~i = 1, 2, 3, ~~~~ 
 \phi_1^{k~sz}\equiv -h3^{k}_{s,z},~~\phi_2^{k~sz}\equiv -h2^{k}_{s,z},~~\phi_3^{k~sz}\equiv -h1^{k}_{s,z}, \ee
for the normal embedding and $\phi_i^{k~sz}$ are listed in Table II. For the flipped embedding, $ \phi_1^{k~sz} $ is changed as $\phi_1^{k~sz}\equiv -h3'^{k}_{s,z}$.

Furthermore, gaugino mass ratios can be read off from Table II since gauginos belong to the multiplets same as gauge bosons in SUSY. That is, for the flipped embedding,
\be M_3:M_2:M_1= h1^{k}_{s,z}:h2^{k}_{s,z}:h3'^{k}_{s,z}, \ee
where $M_a$ are the gaugino masses and $a=3,2,1$ represents the SM SU(3)$\times$SU(2)$\times$U(1) generators. The results are agreed with those in Table V of the paper by S.P. Martin in ref.~\cite{sp}.

The results on $\Phi^{k}_{s,z}$ have more usefulness in examination of GUT. For example, the averaged squared mass of the non-$G_{321}$ singlet gauge bosons (usually called "superheavy" gauge bosons) is given by
\be
\overline{m^2}_{ab}=\sum_i \frac{C(r_i)}{66}g_G^2v_i^2\,\delta_{ab}\equiv M_{\rm HG~ b}^{2}\,\delta_{ab}~~~\text{for}~a,b=13,\,\ldots,78~,  \ee
where the sum runs over, in additional to the non-$G_{321}$ singlet Higgs contained in Eq. (\ref{dim5}), all other Higgs multiplets necessary to realize the gauge symmetry breaking chain in a specific model,
assuming one-step breaking of the grand unified gauge group $E_6$ to the standard model $G_{321}$ at the unification scale $M_{G}$ for simplicity.

In summary, we have investigated effects of dimension-5 operators which are singlets of the grand unified gauge group $E_6$. Considering the breaking chains $E_{6}\mapsto H=SO(10)\times U_{V'}(1)\mapsto SU(5)\times U_{V}(1)\times U_{V'}(1)\mapsto SU(3)\times SU(2)\times U_{Z}(1)\times U_{V}(1)\times U_{V'}(1)$ and $E_{6}\mapsto H=SO(10)\times U_{V'}(1)\mapsto SU(4)\times SU_{L}(2)\times SU_{R}(2)\times U_{V'}(1)\mapsto SU(3)\times SU_{L}(2)\times SU_{R}(2)\times U_{S}(1)\times U_{V'}(1)$, we have derived and given all of the Clebsch-Gordan coefficients $\Phi^{(r)}_{s,z}$ associated with $E_6$ breaking to the standard model, in different bases $\{s,z\}$, upon a uniform absolute normalization for different irreducible representations $r$ contained in the symmetric product of two adjoint reps of $E_6$. We have also discussed some applications of these results to $E_6$ GUT and supersymmetric $E_6$ GUT.

\section*{Acknowledgments}

This research was supported in part by the Natural Science Foundation of
China under grant numbers 10075069, 11375248.

\newpage

\begin{table}[htb]
\begin{center}
\begin{tabular}{ccccccccc}
\hline
$E_6$& r &~ 27 &~ 78 &~ 650 &~ 2430 \\
\hline
 & $ C(r)$ &~ 3 &~ 12 &~ 18 &~ 26   \\
\hline
\hline
$SO(10)$ & r & 10 & 45 & 54 & 210 & 770    \\
\hline
 & $ C(r)$ & 1 & 8 & 10 & 12 & 18    \\
\hline
\end{tabular}
\end{center}
\caption{Quadratic Casimir invariants $C(r)$ for some irreps $r$ of $E_6$ and $SO(10)$, in the conventions explained in the section II of the text. $C({\bf 1})=0$ for the singlet ${\bf 1}$ of any group.\label{C(r)table}}
\end{table}

\begin{table}[htb]
\begin{tabular}{c|c|c||c|c|c|c|c|c|c|c|c|c|c|c|c|c|c|c}
\hline
$E_6$ & SO(10) s & SU(5) z & $h1^{r}_{s,z}$ & $h2^{r}_{s,z}$ & $h3^{r}_{s,z}$ & $h4^{r}_{s,z}$ & $h5^{r}_{s,z}$ & $h6^{r}_{s,z}$ & $h7^{r}_{s,z}$ & $h8^{r}_{s,z}$ & $h9^{r}_{s,z}$ & $h10^{r}_{s,z}$ & $h11^{r}_{s,z}$ & $h12^{r}_{s,z}$ & $h13^{r}_{s,z}$ & $h14^{r}_{s,z}$ & $h15^{r}_{s,z}$ & $N^r_{s,z}$  \\
\hline
\hline
${\bf 1}$ & ${\bf 1}$ 1 & ${\bf 1}$  1 &  1 & 1 & 1 & 1 & 1&  1& 1& 1 & 1 & 1 & 1 & 1&  1& 1& 1 &$1/\sqrt{78}$ \\
\hline
\hline
${\bf 650}$ & ${\bf 1}$ 1& ${\bf 1}$  1 &  1 & 1 & 1 & 1 & 1&  1& 1& 1 &  -5 & -$\frac{5}{4}$ & -$\frac{5}{4}$ & -$\frac{5}{4}$&  -$\frac{5}{4}$& -$\frac{5}{4}$& -$\frac{5}{4}$ & $\frac{1}{2 \sqrt{30}}$ \\
  & ${\bf 45}$ 2& ${\bf 1}$ 1 & 0 & 0 & 0 & 0 & 0 & 0 & 0 & 0 &  0 & $\frac{5}{12}$ & $\frac{5}{12}$ & $\frac{5}{12}$ & -$\frac{5}{4}$ & -$\frac{5}{4}$ & $\frac{25}{12}$ & $\frac{3 \sqrt{3/10}}{10}$ \\
       &    & ${\bf 24}$ 2 & 0 & 0 & 0 & 0 & 0 & 0 & 0 & 0 &  0 & $\frac{5}{12}$ & -$\frac{5}{3}$ & $\frac{5}{2}$ & $\frac{5}{6}$ & -$\frac{5}{4}$ & $0$ & $\frac{3}{10 \sqrt{5}}$ \\
  & ${\bf 54}$ 3& ${\bf 24}$ 1 & 1 & -3/2 & -1/2 & -1/4 & $0$ & -1/4 & 1 & -3/2 & $0$ & $0$ & $0$ & $0$ & $0$ & $0$ & $0$ & $\frac{1}{\sqrt{30}}$\\
  & ${\bf 210}$ 4& ${\bf 1}$ 1 & 1 & 1 & 1 & 1 & -4 & -1& -1& -1 & $0$ & 1/2 & 1/2 & 1/2 & -1/2 & -1/2 & -5/2 & $\frac{1}{4 \sqrt{5}}$\\
        &     & ${\bf 24}$ 2 & 1 & -3/2 & -1/2 & -1/4 & $0$ & 1/4 & -1 & 3/2 & $0$ & -$\frac{1}{8}$ & $\frac{1}{2}$ & -$\frac{3}{4}$ & $\frac{3}{4}$ & -$\frac{9}{8}$ & $0$ & $\frac{1}{3 \sqrt{5}}$\\
        &    & ${\bf 75}$ 3 & 1& 3& -5& -1& $0$ & 1& -1& -3 & $0$ & 1& -1 & -3 & $0$ & $0$ & $0$ & $\frac{1}{12}$\\
\hline
\hline
${\bf 2430}$ & ${\bf 1}$ 1& ${\bf 1}$  1 &  1 & 1 & 1 & 1 & 1&  1& 1& 1 &  27 & -$\frac{9}{4}$ & -$\frac{9}{4}$ & -$\frac{9}{4}$&  -$\frac{9}{4}$ & -$\frac{9}{4}$& -$\frac{9}{4}$ & $\frac{1}{6 \sqrt{26}}$ \\
 & ${\bf 45}$ 2& ${\bf 1}$ 1 &  0 & 0 & 0 & 0 & 0 & 0 & 0 & 0 &  0 & -$\frac{5}{36}$ & -$\frac{5}{36}$ & -$\frac{5}{36}$ & $\frac{5}{12}$ & $\frac{5}{12}$ & -$\frac{25}{36}$ & $\frac{3}{10 \sqrt{2}}$ \\
        &   & ${\bf 24}$ 2 & 0 & 0 & 0 & 0 & 0 & 0 & 0 & 0 &  0 & -$\frac{5}{36}$ & $\frac{5}{9}$ & -$\frac{5}{6}$ & -$\frac{5}{18}$ & $\frac{5}{12}$ & $0$ & $\frac{3 \sqrt{3/5}}{10}$  \\
 & ${\bf 210}$ 3& ${\bf 1}$ 1 & 1 & 1 & 1 & 1 & -4 & -1& -1& -1 & $0$ & -$\frac{3}{2}$ & -$\frac{3}{2}$ & -$\frac{3}{2}$ & $\frac{3}{2}$& $\frac{3}{2}$ & $\frac{15}{2}$ & $\frac{1}{4 \sqrt{15}}$ \\
          &  & ${\bf 24}$ 2 & 1 & -3/2 & -1/2 & -1/4 & $0$ & 1/4 & -1 & 3/2 & $0$ &  $\frac{3}{8}$ & -$\frac{3}{2}$ & $\frac{9}{4}$ & -$\frac{9}{4}$ & $\frac{27}{8}$ & $0$ & $\frac{1}{3 \sqrt{15}}$ \\
          &  & ${\bf 75}$ 3 & 1& 3& -5& -1& $0$ & 1& -1& -3 & $0$ & -3& 3 & 9 & $0$ & $0$ & $0$ & $\frac{1}{12 \sqrt{3}}$ \\
  &  ${\bf 770}$ 4& ${\bf 1}$ 1 &  1 & 1 & 1 & 1 & 16 & -2 & -2& -2 & $0$ & $0$ & $0$ & $0$ & $0$ & $0$ & $0$ & $\frac{1}{6 \sqrt{10}}$\\
         &   & ${\bf 24}~2$ & 1 & -3/2 & -1/2 & -1/4 & $0$ & 7/4 & -7 & 21/2 & $0$ & $0$ & $0$ & $0$ & $0$ & $0$ & $0$ & $\frac{1}{3 \sqrt{210}}$ \\
         &   & ${\bf 75}~3$ & 1& 3& -5& -1& $0$ & -2& 2 & 6 & $0$ & $0$ & $0$ & $0$ & $0$ & $0$ & $0$ & $\frac{1}{6 \sqrt{6}}$ \\
        &    & ${\bf 200}~4$ & 1& 2& 10& -2& $0$ & $0$ & $0$ & $0$ & $0$ & $0$ & $0$ & $0$ & $0$ & $0$ & $0$ & $\frac{1}{2 \sqrt{42}}$\\
\hline
\end{tabular}

\caption{The diagonal part in Eq. (\ref{phir}) of the standard model singlets $\Phi_{s,z}^r$ in each of the irreps $r$ (see Eq. (\ref{reps})) of $E_6$, classified according to their transformation properties under the $SU(5)\subset SO(10) \subset E_6 $ subgroup (SO(10):second column; SU(5): third column  ), in the explicit version with the conventions described in the section II of the text, for the normal embedding. They agree between the normal and the flipped embedding of $G_{321}\subset E_6$, except for the $h3_{s,z}^r$, $h5_{s,z}^r$, $h9_{s,z}^r$. The entries $h3_{s,z}^{\prime r}$, $h5_{s,z}^{\prime r}$, $h9_{s,z}^{\prime r}$ for the flipped embedding are listed in Table III. $N^r_{s,z}$ is the normalization constant which makes $Tr(\Phi^{r}_{s,z} \Phi^{r}_{s,z})=1$ for each irrep r with specific s and z. (For both the normal and the flipped embedding, the $SU(5)$ here contains $SU(3)_C\times SU(2)_L$.)\label{su5}}
\end{table}

\begin{table}[htb]
\begin{tabular}{c||c|c||c|c|c||c|c|c|c|c|c|c|c|c|c|c|c|c}
\hline
$E_6$ & SO(10) s & SU(5) z & $h35^{r}_{s,z}$ & $h39^{r}_{s,z}$ & $h59^{r}_{s,z}$ & $h3^{\prime r}_{s,z}$ & $h5^{\prime r}_{s,z}$ & $h9^{\prime r}_{s,z}$ & $h35^{\prime r}_{s,z}$ & $h39^{\prime r}_{s,z}$ & $h59^{\prime r}_{s,z}$  \\
\hline
\hline
${\bf 1}$ & ${\bf 1}$ 1 & ${\bf 1}$  1 & 0& 0&0 &0& 0&0 &0& 0&0  \\
\hline
\hline
${\bf 650}$ & ${\bf 1}$ 1& ${\bf 1}$  1 & 0& 0&0 & -22/5 & 31/40 & 5/8 & $\frac{9 \sqrt{3/2}}{10}$ & $\frac{-9}{2 \sqrt{10}}$ & $\frac{3 \sqrt{3/5}}{8}$ \\
  & ${\bf 45}$ 2& ${\bf 1}$ 1 & 0 & 0 & $\frac{5 \sqrt{5/3}}{3}$ & 1 & $\frac{1}{24}$ & $\frac{-25}{24}$ & $\frac{-1}{2 \sqrt{6}}$ & $\frac{-7 \sqrt{5/2}}{6}$ & $\frac{7 \sqrt{5/3}}{24}$ \\

       &    & ${\bf 24}$ 2 & 0 & $\frac{-5 \sqrt{5/2}}{3}$ & 0 & 1 & -1 & 0 & $\frac{23}{(4 \sqrt{6}}$ & $\frac{\sqrt{5/2}}{12}$ & $\frac{ \sqrt{5/3}}{2}$  \\
  & ${\bf 54}$ 3& ${\bf 24}$ 1 & -$\sqrt{3/2}$ & 0& 0 & 1/10 & -3/5 & 0 & $\frac{3 \sqrt{3/2}}{20}$ & $\frac{-3}{4 \sqrt{10}}$ & $\frac{-3 \sqrt{3/5}}{2}$
 \\
  & ${\bf 210}$ 4& ${\bf 1}$ 1 &  0& 0 & 0 & -1/5 & 19/20 & -15/4 & $\frac{\sqrt{3/2}}{5}$ & $\frac{3}{\sqrt{10}}$ & $\frac{-\sqrt{3/5}}{4}$  \\
        &     & ${\bf 24}$ 2 & $\frac{3 \sqrt{3/2}}{2}$ &  0 & 0 & -1/5 & -3/10 & 0 & $\frac{-17 \sqrt{3/2}}{40}$ & $\frac{9}{8 \sqrt{10}}$ & $\frac{9 \sqrt{3/5}}{4}$  \\
        &    & ${\bf 75}$ 3 &  0& 0 & 0 & -1/5 & -24/5 & 0 & $\frac{-2 \sqrt{6}}{5}$ & 0 & 0 \\
\hline
\hline
${\bf 2430}$ & ${\bf 1}$ 1& ${\bf 1}$  1 & 0 & 0 & 0 & 122/5 & 79/40& 21/8& $\frac{-39 \sqrt{3/2}}{10}$& $\frac{39}{2 \sqrt{10}}$
& $\frac{-13 \sqrt{3/5}}{8}$ \\
 & ${\bf 45}$ 2& ${\bf 1}$ 1 & 0 & 0&  $\frac{5 \sqrt{5/3}}{3}$ & 1& 1/24& -25/24& $\frac{-1}{2 \sqrt{6}}$ & $\frac{-7 \sqrt{5/2}}{6}$ & $\frac{7 \sqrt{5/3}}{24}$ \\
        &   & ${\bf 24}$ 2 & 0 & $\frac{-5 \sqrt{5/2}}{3}$ & 0 & 1 & -1 & 0 & $\frac{23}{4 \sqrt{6}}$ & $\frac{\sqrt{5/2}}{12}$ & $\frac{ \sqrt{5/3}}{2}$  \\
 & ${\bf 210}$ 3& ${\bf 1}$ 1 & 0 & 0 & 0 & -1/5& 19/20& -15/4& $\frac{\sqrt{3/2}}{5}$ & $\frac{3}{\sqrt{10}}$ & $\frac{-\sqrt{3/5}}{4}$ \\
          &  & ${\bf 24}$ 2 & $\frac{3 \sqrt{3/2}}{2}$ & 0 &0 & -1/5 & -3/10 & 0 & $\frac{-17*\sqrt{3/2}}{40}$ &
      $\frac{9}{8*\sqrt{10}}$ & $\frac{9 \sqrt{3/5}}{4}$ \\
          &  & ${\bf 75}$ 3 & 0 & 0 & 0 & -1/5 & -24/5 & 0 & $\frac{-2 \sqrt{6}}{5}$ & 0 & 0 \\
  &  ${\bf 770}$ 4& ${\bf 1}$ 1 &  0 & 0 & 0 & 1 & 1 & 15 & 0 & -6 $\sqrt{2/5}$ & $\sqrt{3/5}$ \\
         &   & ${\bf 24}~2$ & -21 $\sqrt{3/2}$ & 0 & 0 & 5/2 & -3 & 0 & $\frac{19 \sqrt{3/2}}{4}$ & $\frac{-63}{4 \sqrt{10}}$ & $\frac{-63 \sqrt{3/5}}{2}$ \\
         &   & ${\bf 75}~3$ &  0 & 0 & 0 & -1/5& -24/5& 0& $\frac{-2 \sqrt{6}}{5}$& 0& 0 \\
        &    & ${\bf 200}~4$ & 0 & 0 & 0 & 2/5 & 48/5 &  0 &  $\frac{4 \sqrt{6}}{5}$ &  0 &  0 \\
\hline
\end{tabular}

\caption{The non-diagonal part, i.e., hij, in Eq. (\ref{phir}) of the standard model singlets $\Phi_{s,z}^r$ in each of the irreps $r$ (see Eq. (\ref{reps})) of $E_6$, classified according to their transformation properties under the $SU(5)\subset SO(10) \subset E_6 $ subgroup (SO(10):second column; SU(5): third column), for both the normal and flipped embedding. The entries $h3_{s,z}^{\prime r}$, $h5_{s,z}^{\prime r}$, $h9_{s,z}^{\prime r}$ for the flipped embedding are also listed here. \label{su5n}}

\end{table}

\begin{table}[htb]
\begin{tabular}{c|c|c||c|c|c|c|c|c|c|c|c|c|c|c|c|c|c|c}
\hline
$E_6$ & SO(10) s & $H_1$
 z & $h1^{r}_{s,z}$ & $h2^{r}_{s,z}$ & $h3^{r}_{s,z}$ & $h4^{r}_{s,z}$ & $h5^{r}_{s,z}$ & $h6^{r}_{s,z}$ & $h7^{r}_{s,z}$ & $h8^{r}_{s,z}$ & $h9^{r}_{s,z}$ & $h10^{r}_{s,z}$ & $h11^{r}_{s,z}$ & $h12^{r}_{s,z}$ & $h13^{r}_{s,z}$ & $h14^{r}_{s,z}$ & $h15^{r}_{s,z}$ & $N^r_{s,z}$  \\
\hline
\hline
${\bf 1}$ & ${\bf 1}$ 1 & ${\bf 1}$  1 &  1 & 1 & 1 & 1 & 1&  1& 1& 1 & 1 & 1 & 1 & 1&  1& 1& 1 & $1/\sqrt{78}$ \\
\hline
\hline
${\bf 650}$ & ${\bf 1}$ 1& ${\bf (1,1) }$  1 &  1 & 1 & 1 & 1 & 1&  1& 1& 1 &  -5 & -$\frac{5}{4}$ & -$\frac{5}{4}$ & -$\frac{5}{4}$&  -$\frac{5}{4}$& -$\frac{5}{4}$& -$\frac{5}{4}$ & $\frac{1}{2 \sqrt{30}}$ \\
  & ${\bf 45}$ 2& ${\bf (15,1)}$ 1 &  0 & 0 & 0 & 0 & 0 & 0 & 0 & 0 &  0 & $\frac{5}{12}$ & -$\frac{5}{3}$ & $\frac{5}{2}$ & $\frac{5}{6}$ & -$\frac{5}{4}$ & $0$ & $\frac{3}{10 \sqrt{5}}$ \\
  & ${\bf 54}$ 3& ${\bf (1,1)}$ 1 & 1 & -3/2 & -1/2 & -1/4 & $0$ & -1/4 & 1 & -3/2 & $0$ & $0$ & $0$ & $0$ & $0$ & $0$ & $0$ & $\frac{1}{\sqrt{30}}$\\
  & ${\bf 210}$ 4& ${\bf (1,1)}$ 1 & 0 & 1 & -3/5 & 0 & -2/5 & 0 & 0 & -1 & $0$ & 1/4& -1/4& -1/4& -1/4& 1/4& -1/4& $\frac{1}{2 \sqrt{2}}$\\
        &     & ${\bf (15,1)}$ 2 & 1& 0& -4/5& 0& -6/5& 0& -1& 0  & $0$ & $\frac{1}{4}$ & $\frac{1}{4}$ & -$\frac{3}{4}$ & $\frac{1}{4}$ & -$\frac{3}{4}$ & -$\frac{3}{4}$ & $\frac{1}{2 \sqrt{6}}$\\
        &    & ${\bf (15,3)}$ 3 & 0& 0& 1& 5/12& -1& -5/12& 0& 0 & $0$ & 0& 5/24& 5/8& -5/24& 0& -5/8 & $\sqrt{3}/5$\\
\hline
\hline
${\bf 2430}$ & ${\bf 1}$ 1& ${\bf (1,1)}$  1 &  1 & 1 & 1 & 1 & 1&  1& 1& 1 &  27 & -$\frac{9}{4}$ & -$\frac{9}{4}$ & -$\frac{9}{4}$&  -$\frac{9}{4}$ & -$\frac{9}{4}$& -$\frac{9}{4}$ & $\frac{1}{6 \sqrt{26}}$ \\
 & ${\bf 45}$ 2& ${\bf (15,1)}$ 1 &  0 & 0 & 0 & 0 & 0 & 0 & 0 & 0 &  0 & -$\frac{5}{36}$ & $\frac{5}{9}$ & -$\frac{5}{6}$ & -$\frac{5}{18}$ & $\frac{5}{12}$ & $0$ & $\frac{3 \sqrt{3/5}}{10}$  \\
 & ${\bf 210}$ 3& ${\bf (1,1)}$ 1 & 0& 1& -3/5& 0& -2/5& 0& 0& -1 & $0$ & -3/4& 3/4& 3/4& 3/4& -3/4& 3/4 & $\frac{1}{2 \sqrt{6}}$ \\
          &  & ${\bf (15,1)}$ 2 & 1&0&-4/5&0&-6/5&0&-1&0 & $0$ & -3/4&-3/4&9/4&-3/4& 9/4&9/4 & $\frac{1}{6 \sqrt{2}}$ \\
          &  & ${\bf (15,3)}$ 3 & 0& 0& 1& 5/12& -1& -5/12& 0& 0 & $0$ & 0& -5/8& -15/8& 5/8& 0& 15/8 & $1/5$ \\
  &  ${\bf 770}$ 4& ${\bf (1,1)}$ 1 & 1& 5/2& 19/10& -5/4& 8/5& -5/4& 1& 5/2  & $0$ & $0$ & $0$ & $0$ & $0$ & $0$ & $0$ & $\frac{1}{3 \sqrt{10}}$\\         &   & ${\bf (15,1)}~2$ & 0& 0& 1& 0& 2/3& 0& 0& -5/6 & $0$ & $0$ & $0$ & $0$ & $0$ & $0$ & $0$ & $\frac{\sqrt{6}}{5}$ \\
         &   & ${\bf (15,3)}~3$ & 0& 0& 1& -5/24& -1& 5/24& 0& 0& $0$ & $0$ & $0$ & $0$ & $0$ & $0$ & $0$ & $\frac{ 2 \sqrt{2}}{5}$ \\
        &    & ${\bf (84,1)}~4$ & 1& 0& 32/5& 0& 48/5& 0& -4& 0 & $0$ & $0$ & $0$ & $0$ & $0$ & $0$ & $0$ & $\frac{1}{6 \sqrt{10}}$\\
\hline
\end{tabular}

\caption{The diagonal part in Eq. (\ref{phir}) of the standard model singlets $\Phi_{s,z}^r$ in each of the irreps $r$ (see Eq. (\ref{reps})) of $E_6$, classified according to their transformation properties under the $H_1=SU(4)\times SU_R(2)\subset SO(10) \subset E_6 $ subgroup (SO(10):second column; $SU(4)\times SU_R(2)$: third column, all entries are $SU_{L}(2)$ singlet), in the explicit version with the conventions described in the section II of the text, for the normal embedding. They agree between the normal and the flipped embedding of $G_{321}\subset E_6$, except for the $h3_{s,z}^r$, $h5_{s,z}^r$, $h9_{s,z}^r$. The entries $h3_{s,z}^{\prime r}$, $h5_{s,z}^{\prime r}$, $h9_{s,z}^{\prime r}$ for the flipped embedding are listed in Table V. $N^r_{s,z}$ is the normalization constant which makes $Tr(\Phi^{r}_{s,z} \Phi^{r}_{s,z})=1$ for each irrep r with specific s and z. (For both the normal and the flipped embedding, the $SU(4)$ here contains $SU(3)_C$.)\label{su4}}
\end{table}

\begin{table}[htb]
\begin{tabular}{c||c|c||c|c|c||c|c|c|c|c|c|c|c|c|c|c|c|c}
\hline
$E_6$ & SO(10) s & $SU(4)\times SU_R(2)$ z & $h35^{r}_{s,z}$ & $h39^{r}_{s,z}$ & $h59^{r}_{s,z}$ & $h3^{\prime r}_{s,z}$ & $h5^{\prime r}_{s,z}$ & $h9^{\prime r}_{s,z}$ & $h35^{\prime r}_{s,z}$ & $h39^{\prime r}_{s,z}$ & $h59^{\prime r}_{s,z}$  \\
\hline
\hline
${\bf 1}$ & ${\bf 1}$ 1 & ${\bf (1,1)}$  1 & 0& 0&0 &0& 0&0 &0& 0&0  \\
\hline
\hline
${\bf 650}$ & ${\bf 1}$ 1& ${\bf (1,1)}$  1 & 0& 0&0 & -22/5 & 31/40 & 5/8 & $\frac{9 \sqrt{3/2}}{10}$ & $\frac{-9}{2 \sqrt{10}}$ & $\frac{3 \sqrt{3/5}}{8}$ \\
  & ${\bf 45}$ 2& ${\bf (15,1)}$ 1 & 0 & $\frac{-5 \sqrt{5/2}}{3}$ & 0 & 1 & -1 & 0 & $\frac{23}{(4 \sqrt{6}}$ & $\frac{\sqrt{5/2}}{12}$ & $\frac{ \sqrt{5/3}}{2}$  \\
  & ${\bf 54}$ 3& ${\bf (1,1)}$ 1 & -$\sqrt{3/2}$ & 0& 0 & 1/10 & -3/5 & 0 & $\frac{3 \sqrt{3/2}}{20}$ & $\frac{-3}{4 \sqrt{10}}$ & $\frac{-3 \sqrt{3/5}}{2}$
 \\
  & ${\bf 210}$ 4& ${\bf (1,1)}$ 1 & -$\sqrt{6}/5$ & 0 & 0 & 0 & -$\frac{5}{8}$ & -$\frac{3}{8}$ & 0 & 0 & -$\frac{15}{8}$  \\
        &     & ${\bf (15,1)}$ 2 & $\frac{2 \sqrt{6}}{5}$ &  0 & 0 & -1/5 & -27/40 & -9/8 & $\frac{-3 \sqrt{3/2}}{10}$ & $\frac{3}{2 \sqrt{10}}$ & $\frac{9 \sqrt{3/5}}{8}$  \\
        &    & ${\bf (15,3)}$ 3 & -$\frac{1}{2\sqrt{6}}$ &  0 & 0 & 0 & 15/16 & -15/16 & $\frac{5}{8 \sqrt{6}}$ & $\frac{\sqrt{5/2}}{8}$ & -$\frac{\sqrt{15}}{16}$  \\
\hline
\hline
${\bf 2430}$ & ${\bf 1}$ 1& ${\bf (1,1)}$  1 & 0 & 0 & 0 & 122/5 & 79/40& 21/8& $\frac{-39 \sqrt{3/2}}{10}$& $\frac{39}{2 \sqrt{10}}$
& $\frac{-13 \sqrt{3/5}}{8}$ \\
 & ${\bf 45}$ 2& ${\bf (15,1)}$ 1 &  0 & $\frac{-5 \sqrt{5/2}}{3}$ & 0 & 1 & -1 & 0 & $\frac{23}{4 \sqrt{6}}$ & $\frac{\sqrt{5/2}}{12}$ & $\frac{ \sqrt{5/3}}{2}$  \\
 & ${\bf 210}$ 3& ${\bf (1,1)}$ 1 & -$\frac{\sqrt{6}}{5}$ & 0 & 0 & 0& -5/8& -3/8& 0 & 0 & $\frac{-\sqrt{15}}{8}$ \\
          &  & ${\bf (15,1)}$ 2 & $\frac{2 \sqrt{6}}{5}$ & 0 &0 & -1/5 & -27/40 & -9/8 & $\frac{-3 \sqrt{3/2}}{10}$ &
      $\frac{3}{2\sqrt{10}}$ & $\frac{9 \sqrt{3/5}}{8}$ \\
          &  & ${\bf (15,3)}$ 3 & -$\frac{1}{2 \sqrt{6}}$ & 0 & 0 & 0 & 15/16 & -15/16 & $\frac{5}{8 \sqrt{6}}$ & $\frac{\sqrt{5/2}}{8}$ & -$\frac{\sqrt{15}}{16}$ \\
  &  ${\bf 770}$ 4& ${\bf (1,1)}$ 1 & $\frac {3 \sqrt{3/2}}{5}$ & 0 & 0 & 1/10 & 19/10 & 3/2 & $\frac{3 \sqrt{3/2}}{20}$ & -$\frac{3}{4 \sqrt{10}}$ & $\sqrt{3/5}$ \\
         &   & ${\bf (15,1)}~2$ & $\sqrt{2/3}$ & 0 & 0 & 0 & 25/24 & 5/8 & 0& 0& $\frac{5\sqrt{5/3}}{8}$ \\
         &   & ${\bf (15,3)}~3$ &  -$\frac{1}{2\sqrt{6}}$ & 0 & 0 & 0& 15/16& -15/16& $\frac{5}{8\sqrt{6}}$& $\frac{ \sqrt{5/2}}{8}$& -$\frac{\sqrt{15}}{16}$ \\
        &    & ${\bf (84,1)}~4$ & -$\frac{16\sqrt{6}}{5}$ & 0 & 0 & 8/5& 27/5& 9& $\frac{6\sqrt{6}}{5}$& -6 $\sqrt{2/5}$& -9
        $ \sqrt{3/5}$ \\
\hline
\end{tabular}

\caption{The non-diagonal part, i.e., hij, in Eq. (\ref{phir}) of the standard model singlets $\Phi_{s,z}^r$ in each of the irreps $r$ (see Eq. (\ref{reps})) of $E_6$, classified according to their transformation properties under the $SU(4)\times SU_R(2)\subset SO(10) \subset E_6 $ subgroup (SO(10):second column; $SU(4)\times SU_R(2)$: third column), for both the normal and flipped embedding. The entries $h3_{s,z}^{\prime r}$, $h5_{s,z}^{\prime r}$, $h9_{s,z}^{\prime r}$ for the flipped embedding are also listed here. \label{su4n}}

\end{table}

\end{document}